\documentclass[12pt,preprint]{aastex}
\usepackage{epsfig}
\usepackage{graphics,graphicx,amsmath}

\newcommand{\lapprox} {\, \lower3pt\hbox{$\sim$}\llap{\raise2pt\hbox{$<$}}\,}
\newcommand{\gapprox} {\, \lower3pt\hbox{$\sim$}\llap{\raise2pt\hbox{$>$}}\,}

\begin{document}

\title{Return currents and energy transport in the solar flaring atmosphere}
\author{Anna Codispoti\altaffilmark{1},Gabriele Torre\altaffilmark{1,2},Michele Piana\altaffilmark{1,2} \&
Nicola Pinamonti\altaffilmark{1}
}

\altaffiltext{1}{Dipartimento di Matematica, Universit\`a di Genova, via Dodecaneso 35, 16146 Genova, Italy}

\altaffiltext{2}{CNR - SPIN, via Dodecaneso 33, I-16146
Genova, Italy}

\begin{abstract}
According to a standard ohmic perspective, the injection of accelerated electrons into the flaring region violates local charge equilibrium and therefore, in response, return currents are driven by an electric field to equilibrate such charge violation. In this framework, the energy loss rate associated to these local currents has an ohmic nature and significantly shortens the acceleration electron path. In the present paper we adopt a different viewpoint and, specifically, we study the impact of the background drift velocity on the energy loss rate of accelerated electrons in solar flares. We first utilize the Rutherford cross-section to derive the formula of the energy loss rate when the collisional target has a finite temperature and the background instantaneously and coherently moves up to equilibrate the electron injection. We then use the continuity equation for electrons and imaging spectroscopy data provided by {\em{RHESSI}} to validate this model. Specifically, we show that this new formula for the energy loss rate provides a better fit of the experimental data with respect to the model based on the effects of standard ohmic return currents.

%
%
\end{abstract}


\section{Introduction}
In arc-shaped solar flares, huge amounts of electrons are accelerated due to magnetic reconnection. 
%
In the standard picture (\citealt{fle11}) accelerated electrons are injected at the top of the loop-shaped flare, move down along magnetic field lines and, during their motion, loose energy because of various mechanisms. Coulomb collisions with ambient particles represent the most relevant mechanism explaining energy losses for accelerated electrons in loop-shaped flares. In this framework, the predominant energy loss process involves the interaction between the injected electrons and the electrons in the flaring target. Specifically, under the assumption of a cold target, the energy loss rate formula depends as $1/E$ on the electron energy $E$ (\citealt{em78}) while in a warm target a more complicated formula accounts for the temperature $T$ of the ambient medium (\citealt{sp62}, \citealt{lg63}). However, other important effects may impact on this process, such as wave-particles interactions (\citealt{hoyngmelrose77}, \citealt{hannahkontar11}) and return currents. This paper focuses on the role of return currents in the energy loss mechanism during flares and specifically utilizes measurements observed by the {\em{Reuven Ramaty High Energy Solar Spectroscopic Imager (RHESSI)}} to empirically study the mechanism of energy loss in the presence of return currents.

In their pioneering paper, \cite{knightsturrock77} analyzed the effect of return currents on accelerated electron motion by means of electrostatic fields and using Vlasov equation. The combination of reverse current electrostatic field with collisions has been discussed in \cite{emslie80}, where the impact of return currents on hard X-ray emission has been studied as well.
Bounds for the size of the electrostatic term due to unstable electron-ion drifts are found in \cite{emslie81}.
In \cite{brownbingham84}, another approach using electrostatic modeling is presented. 
The influence of inductive fields and their interplay with the electrostatic one have been thoroughly discussed and compared by   
\cite{larosaemslie89} and by \cite{oord90}. 
Starting form observations, \cite{mcclymontcanfield86} argued that the areas of injection are considerably small and thus very large return currents are expected.
%
%
%
More recently \cite{alexanderdaou07} used RHESSI observations to estimate the injection area, confirming the effects on hard X-ray emission predicted in \cite{emslie80}. A complete treatment of the many possible kinetic energy loss mechanisms, including ohmic losses, has been more recently presented in \cite{zharkovagordovskyy05}.

%
%
%
%
%
%
%
%
%
%
%

All previous papers are based on an ohmic viewpoint according to which the injection of a large number of accelerated electrons into the loop violates the local charge equilibrium and, in response to this violation, local currents are established by flare background electrons. In this picture, the generated return currents are driven by an electric field and energy losses of ohmic nature shorten the accelerated electron path before thermalization. However a different model describing the restoration of charge equilibrium is possible, in which the production of return currents occurs instantaneously, the background particles moves up coherently along the field line, while the energy loss rate due to Coulomb collisions is modified in a very peculiar manner: since the single accelerated electron appears more energetic in the rest frame of the background motion and since the energy loss rate decays with energy, we expect the resulting Coulomb collision energy loss rate to be lower than the one computed for background particles with vanishing drift velocity.
 
%
%


The aim of the present paper is to investigate the impact of this coherent background motion on the energy loss rate for accelerated electrons and to compare it with the competing influence of ohmic forces induced by return currents. This comparison will be performed by following the same empirical approach adopted by \cite{to12}: we will (1) use hard X-ray imaging spectroscopy data observed by {\em{RHESSI}} to reconstruct mean electron flux images of extended sources (\citealt{pianaetal07}); and (2) apply the electron continuity equation to select the model for energy loss rate that best fits the empirical electron maps.\\
 
 The plan of the paper is as follows. In the next section we shall derive the energy loss rate for Coulomb collisions when 
a coherent motion is added to the background Maxwellian distribution. 
In the third section we shall describe the procedure we have adopted to compare the various energy loss rate models.
The comparison of such models with {\em{RHESSI}} observations is presented in the fourth section and finally some conclusions are drawn in the last section.

\section{A general model for the Coulomb collisional energy loss rate}
The main ingredients at the basis of our model are two formulas obtained by \cite{ruth11} and \cite{bubu62}, respectively. In the first one the differential cross section, which 
describes the probability of a single collision of a particle of charge $ze_-$ and mass $m$ 
travelling with a speed  ${\bf v}$ against a target particle of mass $M$ and charge $Ze_-$ moving at velocity ${\bf w}$, is 
\begin{equation}\label{rutherford}
\frac{d\sigma}{d\Omega} = \frac{(Zz)^2e_-^4}{4\left( \frac{M m}{M+m}\right)^2} \frac{1}{|{\bf v}-{\bf w}|^4} \frac{1}{\sin^4\left( \frac{\theta}{2}\right)} \;,\qquad \theta \geq \theta_0\;.
\end{equation}
This expression is non zero only for scattering angles $\theta$ bigger than a fixed small angle $\theta_0$
which is related to the Debye screening length. 

%
The second formula describes the collisional loss rate for a particle traveling in a plasma, whose particle velocities are distributed as 
$f({\bf w})$, and is given by
\begin{equation}\label{dedt2}
\left(\frac{dE}{dt}\right)_c = - 4\pi\frac{(Zz)^2e_-^4 \Lambda}{M}   \int\frac{({\bf v}-{\bf w})\cdot({\bf v}+\frac{M}{m}{\bf w} )}{|{\bf v}-{\bf w}|^3}   f({\bf w})\, d^3{\bf w}\;.
\end{equation}
Here the Coulomb logarithm $\Lambda$ is related to $\theta_0$ by
\begin{equation}\label{clog}
\Lambda =  - \log \left(\sin \frac{\theta_0}{2}\right)\simeq - \log \frac{\theta_0}{2}\;.
\end{equation}

In order to discuss the role of a coherent motion in the energy loss rate, we shall follow the derivation of (\ref{dedt2}) given by \cite{bubu62} and modify the form of the target distribution. We shall here assume that the background particles are in a stationary solution of the Boltzmann equation, i.e. that they are distributed according to the Maxwell law but with a coherent motion induced by a drift velocity ${\bf w}_0$, and a thermal velocity ${w}_T $ such that $M w_T^2/2=k_B T$, where $T$ is the temperature of the distribution and $k_B$ is the Boltzmann constant.
Under these hypotheses we have
\begin{equation}\label{fw}
f({\bf w}) = \frac{n}{\pi^{3/2} w_T^3} e^{-\left(\frac{{\bf w}-{\bf w}_0}{{w}_T}\right)^2}\;,
\end{equation}
where $n$ is the target electron density.
Integrating over the target velocities and expressing the result in terms of ${\bf v}'={\bf v}-{\bf w}_0$ we get 
\begin{multline}\label{questa}
\left(\frac{dE}{dt}\right)_c = 
-  
\frac{2K n}{M\; v'} 
\left[
\left(
1
+ 
\left(1 + \frac{M}{m}\right) 
\frac{{\bf v}'\cdot {\bf w}_{0}}{{v'}^2} 
\right)
\right.
\text{erf} \left( \frac{v'}{w_T} \right)
\\
- 
\left.
\frac{2}{\sqrt{\pi}}\left(1+\frac{M}{m}\right)
\left(
1
+\frac{{\bf v}'\cdot {\bf w}_{0}}{{v'}^2}
\right)
\frac{v'}{w_T}
e^{-\left(\frac{v'}{w_T}\right)^2 }
\right]
\;,
\end{multline}
where, $K= 2\pi (Zz)^2  e_-^4  \Lambda$ and $\text{erf}$ is the standard error function.
It is possible to point out two terms in equation \eqref{questa} characterized by two different physical meanings. The first term corresponds to 
the pure thermal energy loss rate computed in a frame where the coherent velocity of the background vanishes (${\bf w}_{0}=0$). The latter one is proportional to ${\bf v}'\cdot {\bf w}_{0}$ and cannot be directly inferred from the pure thermal energy loss rate. Equation (\ref{questa}) can be simplified by assuming that ${\bf w}_0$ and ${\bf v}$ are anti-parallel and that the electron-electron collisions give the most important contribution to the energy loss. Therefore, if
%
%
%
\begin{equation}\label{assumption}
{\bf v} = v {\bf e}_z   \;,\qquad {\bf w}_0  = - w_0 {\bf e}_z
\end{equation}
and $M=m$, $Z=z=1$, then
\begin{multline}
\left(\frac{dE}{dt}\right)_c = - \frac{2K n}{m\; (v+w_0)} \left[\left(1-2\frac{w_0}{v+w_0} \right)\right.
\text{erf} \left( \frac{v+w_0}{w_T} \right)\\
-\left.\label{eloss}\frac{4}{\sqrt{\pi}}\left(1-\frac{w_0}{v+w_0}\right)\frac{v+w_0}{w_T}e^{-\left(\frac{v+w_0}{w_T}\right)^2 }\right].
\end{multline}
When $w_0$ vanishes the previous expression reduces to the standard energy loss rate formula that can be found in \cite{sp62}.
As previously noticed in the general case, also with the further assumption stated above, the energy loss rate is formed by two contributions. The one which cannot be directly inferred from the thermal energy loss rate formula 
is relevant also in the zero temperature limit ($w_T\to0$) where we obtain
$$
\left(\frac{dE}{dt}\right)_c = 
-\frac{2 K\,n}{m}
\left(
\frac{1}{v+w_0} - 2 \frac{w_0}{{(v+w_0)}^2}
\right)
=
-\frac{2 K\,n }{m}
\frac{v-w_0}{{(v+w_0)}^2}\;.
$$

\begin{center}
\begin{figure}[!h]
\centering
\includegraphics[width=12.cm]{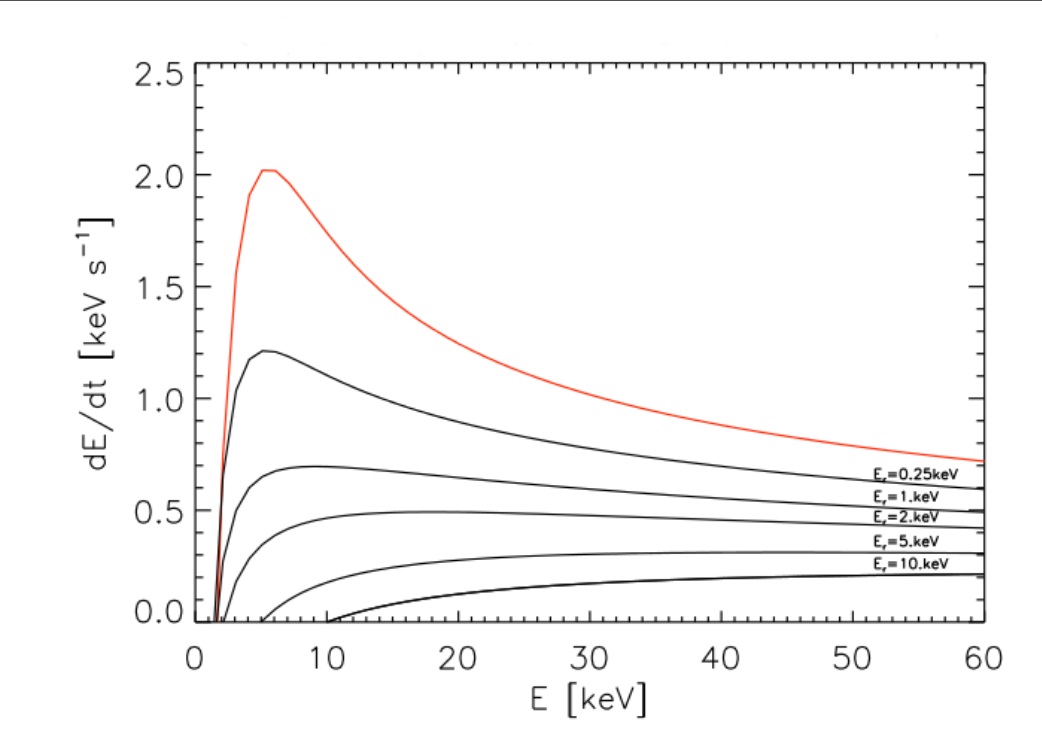}
\caption{The energy loss rate for the electron-electron cross section with thermal energy fixed at $1.6$ keV for various background  
velocities $w_0$. In the plot we have used as labels the kinetic energy $E_0$ associated to the corresponding drift velocity $w_0$. The red line indicates the case when $E_0=0$.
}
\label{fig:eloss-return-currents}
\end{figure}
\end{center}
Figure \ref{fig:eloss-return-currents} provides a qualitative analysis of the energy loss rate (\ref{questa}) for different values of the kinetic energy $E_0$ associated to the drift velocity $w_0$. From this analysis it clearly follows that, when the intensity of the background motion increases, the energy loss rate of the injected electrons decreases. Further, all different forms of the energy loss rate for all different values of $E_0$ assume an asymptotic behavior $1/E$ at high electron energies.

\section{Selection of the energy loss model}

We now validate the model in equation \eqref{eloss}
with electron flux maps  reconstructed from RHESSI data,
following the approach introduced by \cite{to12} .

In the standard flare picture, electrons move from the injection region towards foot-points along magnetic field lines, loosing energy along the paths. If $F(E,s)$  (electrons cm$^{-2}$ s$^{-1}$ keV$^{-1}$) is the electron flux, differential in energy along the direction s and $N(s)$ is the column depth, \cite{ebb01}  proved that $g(s;E) = N(s)F(E,s)$ (electrons cm$^{-4}$ s$^{-1}$ keV$^{-1}$) satisfies the continuity equation
  
\begin{equation}\label{eqn:continuitybeamflux}
\pm\frac{\partial}{\partial s}g(s;E)-\frac{\partial}{\partial E}\left(\left(\frac{dE}{ds}\right)_{tot}g(s;E)\right)=S(s,E),
\end{equation}
where the first sign is positive for electrons moving towards larger $s$ and negative otherwise. In this equation, $g(s;E)$ corresponds to the mean electron flux maps that can be reconstructed from {\em{RHESSI}} visibilities (\citealt{pianaetal07}) while $S(s,E)$ is the source term coding the information on the injection region.
The continuity equation can be interpreted to obtain 
\begin{equation}\label{eqn:continuity_definitivo}
R(s,E)=
-\left(\frac{dE}{ds}\right)_{tot}+ \frac{1}{g(s;E)}\int_E^\infty  
S(s;E')dE',
\end{equation}
where 
\begin{equation}\label{eqn:expcontr}
R(s,E):=\pm\frac{1}{g(s;E)}\int_E^\infty \frac{\partial g(s;E')}{\partial s}dE'
\end{equation}
is an empirical quantity determined from the electron maps. In \eqref{eqn:continuitybeamflux} and \eqref{eqn:continuity_definitivo}-\eqref{eqn:expcontr} the source term $S(s,E)$ describes the injection of electrons in the flare region and thus it is an energy gain term. Following \cite{guo12} here we assume
\begin{equation}\label{eqn:iniezione}
S(E,s)=\begin{cases} {h_s}\left(\frac{E}{E_s}\right)^{-\delta};\qquad & |s|\leq\frac{L}{2}\\
0;\qquad & |s|>\frac{L}{2}
\end{cases}
\end{equation}
where $h_s$ (electrons cm$^{-5}$ keV$^{-1}$ s$^{-1}$) is the source amplitude averaged along the line of sight, $\delta$ is the spectral index of the injected electrons, $L$ represents the length of the injected region 
and $E_s$ is fixed equal to 10 keV. Since here we are interested in the energy domain, and since we shall average over different $s$ as in \cite{to12}, the choice of a box shaped injection as in (\ref{eqn:iniezione}) is not really restrictive.\\

The total energy loss rate 
$$
\left(\frac{dE}{ds}\right)_{tot} = \frac{1}{v}\left(\frac{dE}{dt}\right)_{tot} 
$$ 
in \eqref{eqn:continuity_definitivo} represents the core of the present analysis. In the following we will consider four possible situations:
\begin{itemize}
\item {\bf Model 1}: hot target with charge equilibrium. This is the case pictured by \eqref{eloss}, in which the collisional target has a finite temperature and the background instantaneously reacts to the electron injection by means of a coherent motion.
\item {\bf Model 2}: cold target with charge equilibrium. This is what happens when in \eqref{eloss} one fixes $E_T=\frac{1}{2}mw_T^2=0$.
\item {\bf Model 3}: ohmic losses. In this case there is no drift velocity in the collisional term \eqref{eloss} and the energy loss rate has the form
\begin{equation}
\left(\frac{dE}{ds}\right)_{tot} = \left(\frac{dE}{ds}\right)_{c} + \left(\frac{dE}{ds}\right)_{ohm} 
\end{equation}
where the second term on the r.h.s. describing the ohmic losses is given by
\begin{equation}
\left(\frac{dE}{ds}\right)_{ohm} = -e_-\mathcal{E} = -ne_-^2\eta w_0.
\end{equation}
Here ${\mathcal E}$ is the electric field which drives the return currents, $\eta$ is the resistivity, $n$ is the background density and $w_0$ is again the background velocity. The first term at the r.h.s. is the standard energy loss formula that can be found in \cite{sp62}, corresponding to the case when no background velocity is present and the target temperature is finite.
\item {\bf Model 4}: hot target without return currents. This is the model described by \cite{sp62} and already discussed by \cite{to12}.
\end{itemize}
For all four models the number of free parameters to fit against the empirical $R(E,s)$ is always two, and in all cases these parameters are the target density $n$ and the averaged source amplitude $h_s$ in \eqref{eqn:iniezione}. In fact, the target temperature can be fitted by using spatially integrated spectroscopy. This same spectroscopy and the charge equilibrium constraint allow  fixing the drift velocity $w_0$. More specifically, assuming that the injected electron flux
\begin{equation}\label{h0}
\int_{E_s}^{\infty}\int_{-\frac{L}{2}}^{\frac{L}{2}}h_s \left(\frac{E_s}{E}\right)^{\delta} dE ds = h_s  \frac{E_s L} {\delta-1},
\end{equation}
is equal to the flux $nN\sqrt{\frac{2}{m} E_0}$ associated to the return currents leads to 
\begin{equation}\label{equilibrium-constraint}
h_s  \frac{E_s L} {\delta-1}=\frac{EM}{A}\sqrt{\frac{2}{m} E_0}.
\end{equation}
Since the emission measure $EM$ can be inferred from spectroscopy and the flare area $A$ from the reconstructed electron maps, equation \eqref{equilibrium-constraint} represents a constraint for $E_0$ and therefore for $w_0$.

\section{Models versus observations}
In this section, we present the results of the analysis stated above for 
various time bins of three different flares, namely those occurred on April 15 2002, April 16 2002
and May 21 2004. The time intervals, and the relevant physical parameters are described in the following Table 1. The temperature $T$, the spectral index $\gamma=\delta-1$\footnote{This relation holds exactly in the case of the Kramers bremsstrahlung cross-section, while for more general formulas it is true just approximately (\citealt{bretal08}). In this context, the accuracy of the results is not affected by the use of this approximation.} and $EM$ are obtained from photon spectral fitting. The flare area $A$ is computed from the reconstructed electron maps. Specifically, we used the map at 14 keV as reference image and summed up the areas of pixels with intensity higher than $10\%$ of the maximum intensity. The spatial region $\Delta s=[s_{min}, s_{max}]$ over which the average of $R(E,s)$ is computed is determined as in \cite{to12} (see Figure 4 of that paper). Specifically, the electron flux images are considered for all energies and for each row of each image we took the pixel with maximum intensity. We thus drew a path in each image approximating a field line. In this path we fixed at $s=0$ the pixel with maximum intensity and assumed that moving toward the right foot-point increases $s$ while moving toward the left foot-point decreases $s$.

\begin{figure}
\begin{tabular}{cc}
\includegraphics[width=0.5\textwidth]{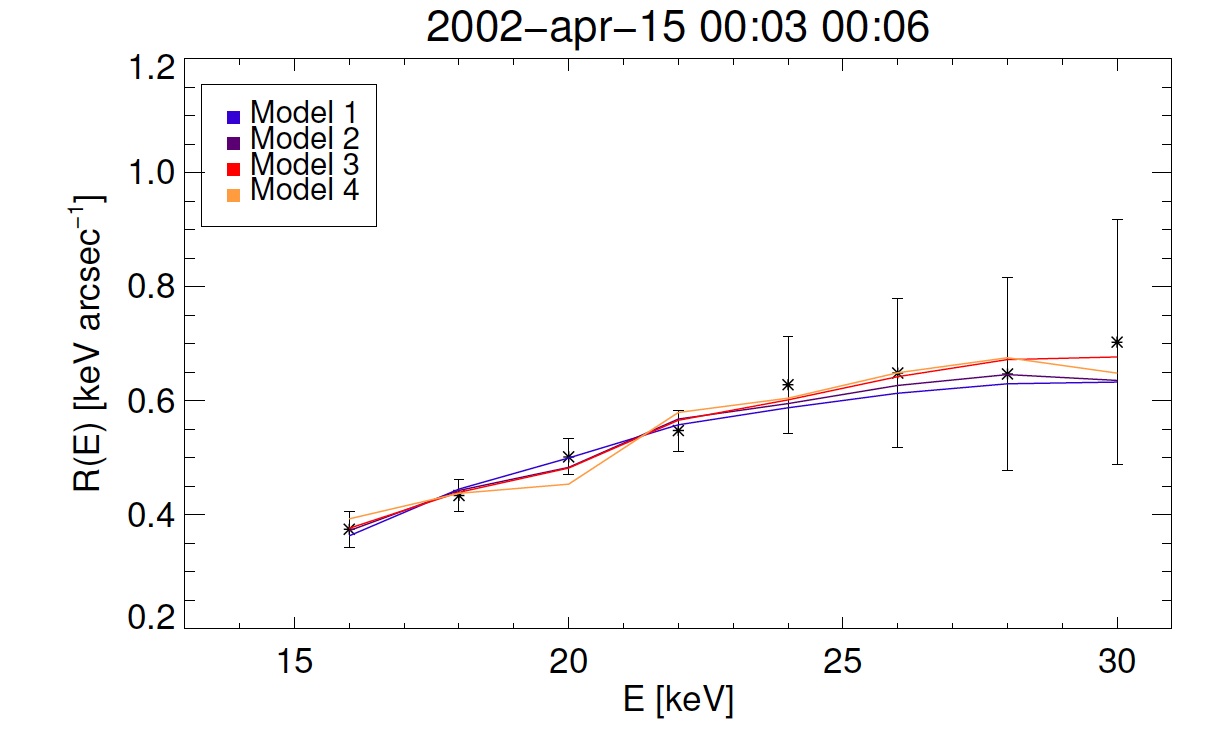}
\includegraphics[width=0.5\textwidth]{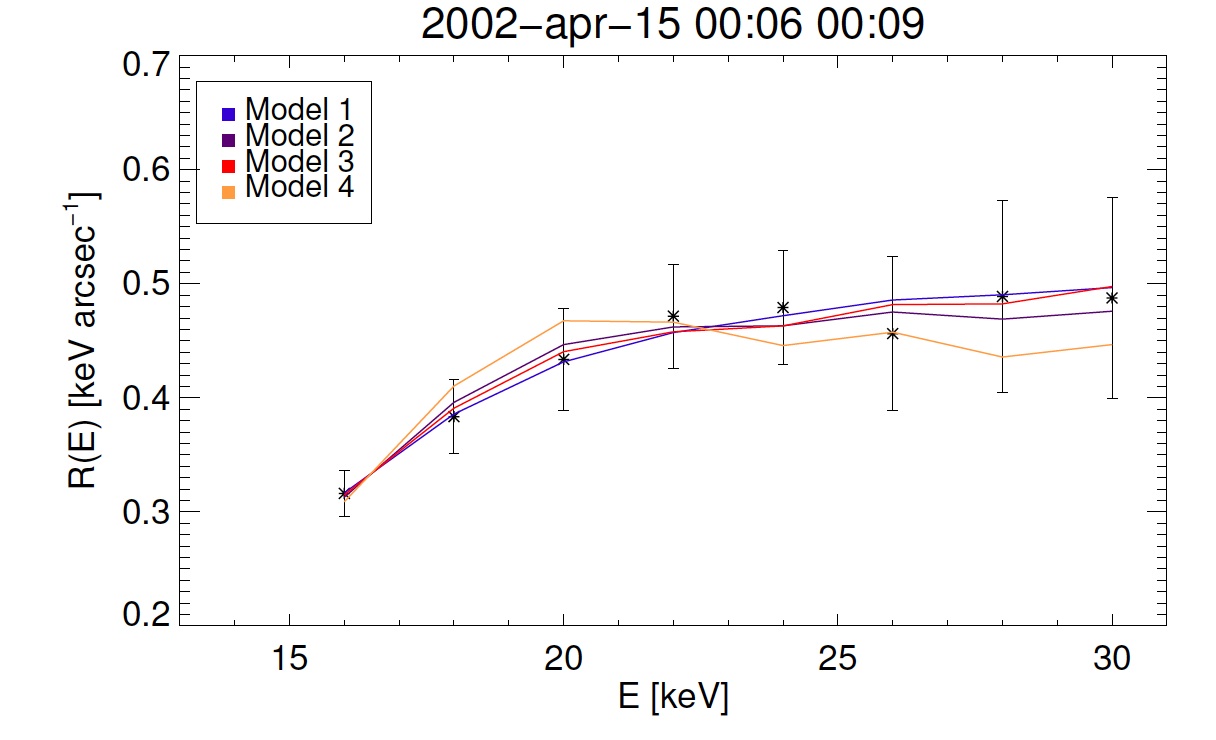}\\
\includegraphics[width=0.5\textwidth]{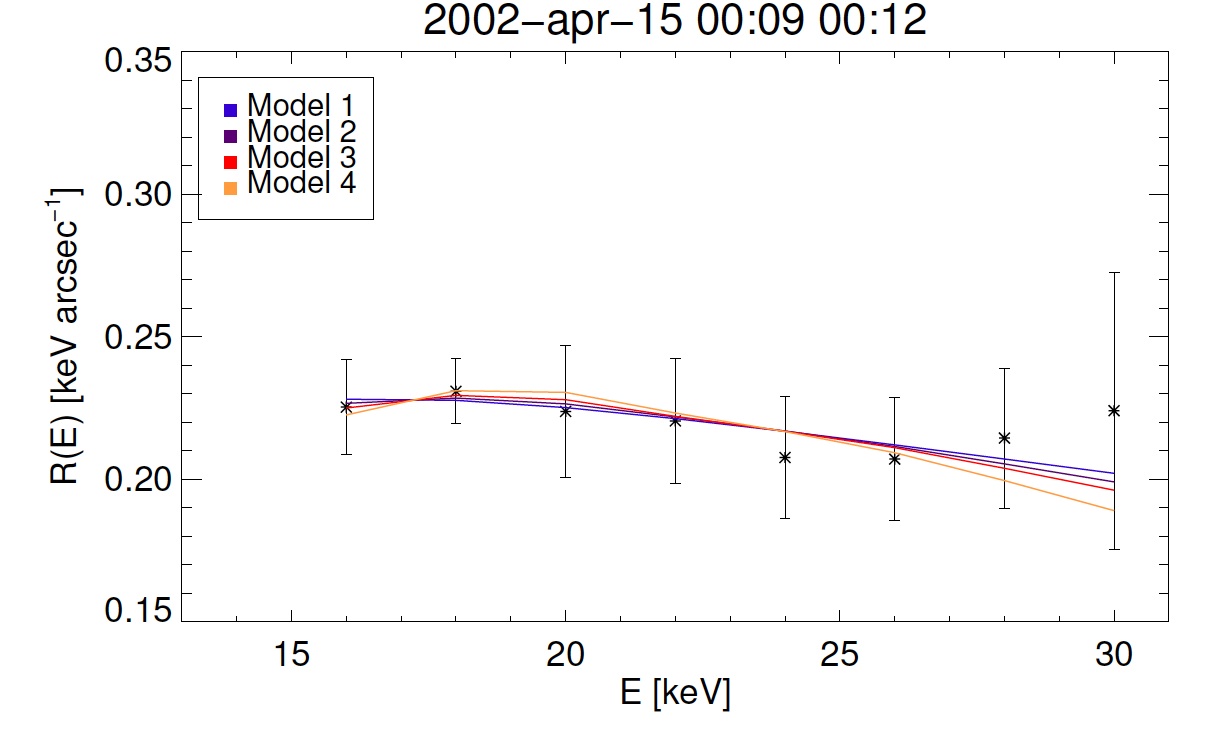}\\
\includegraphics[width=0.5\textwidth]{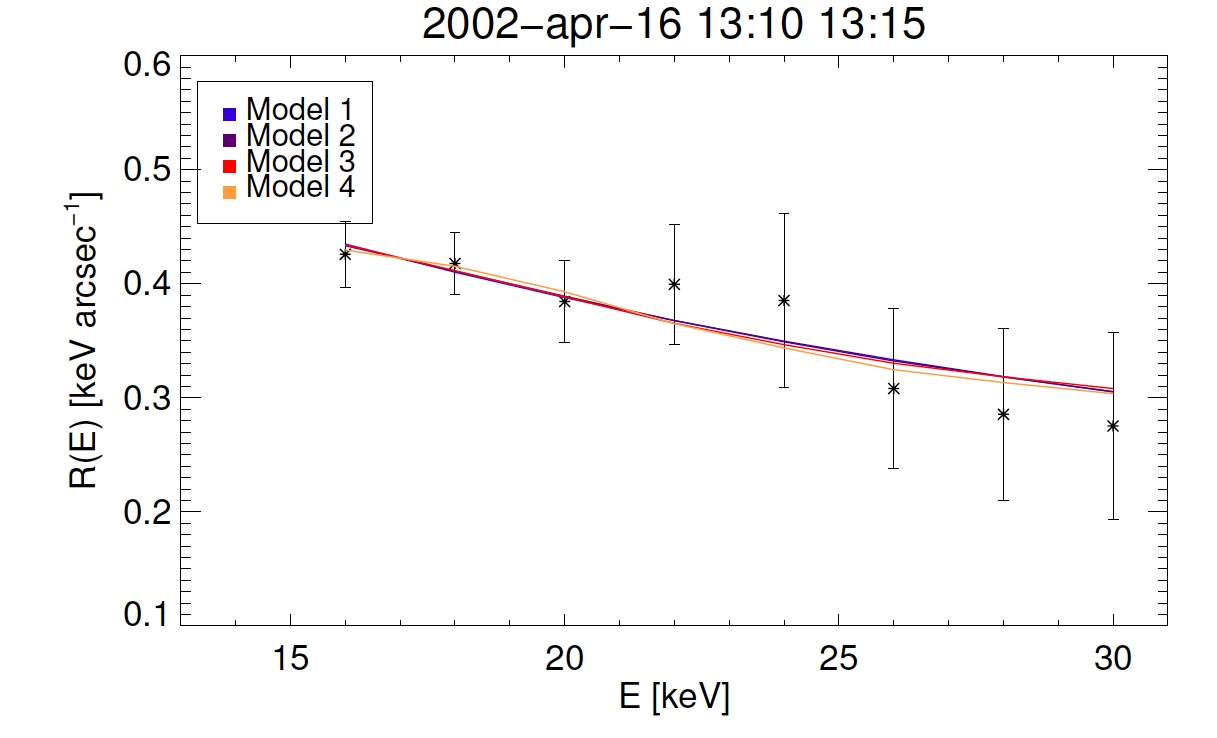}
\includegraphics[width=0.5\textwidth]{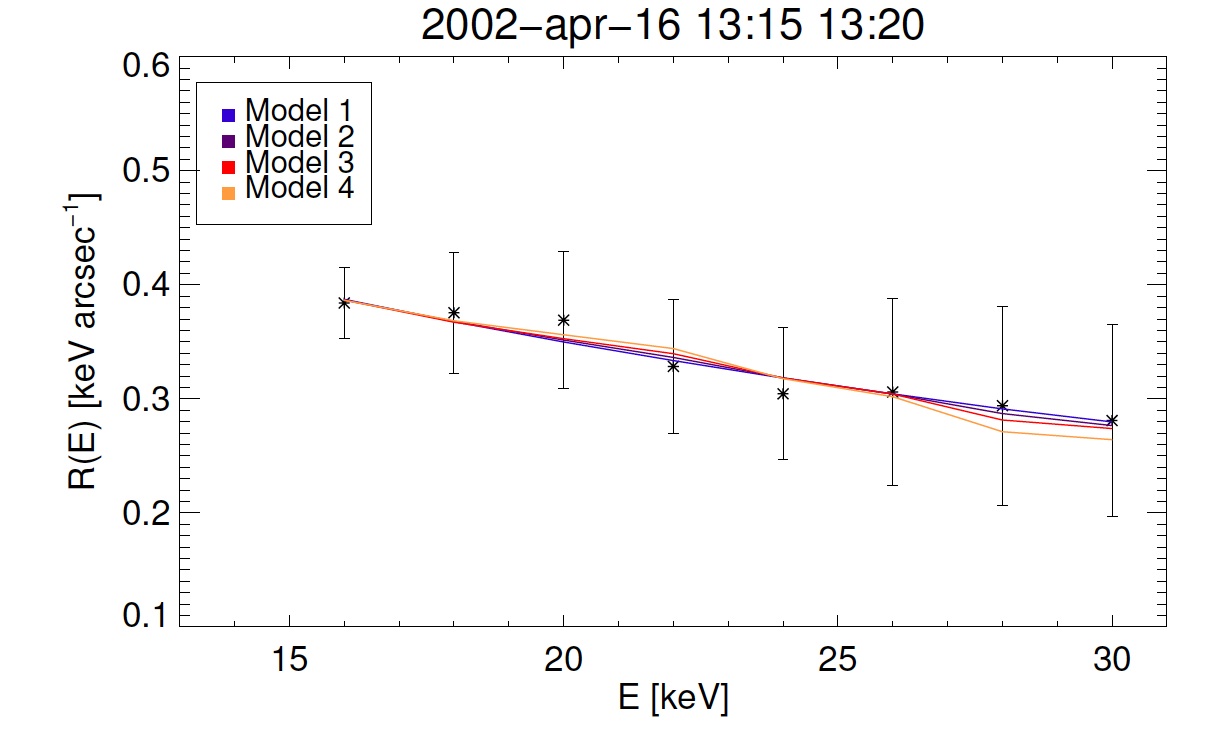}\\
\includegraphics[width=0.5\textwidth]{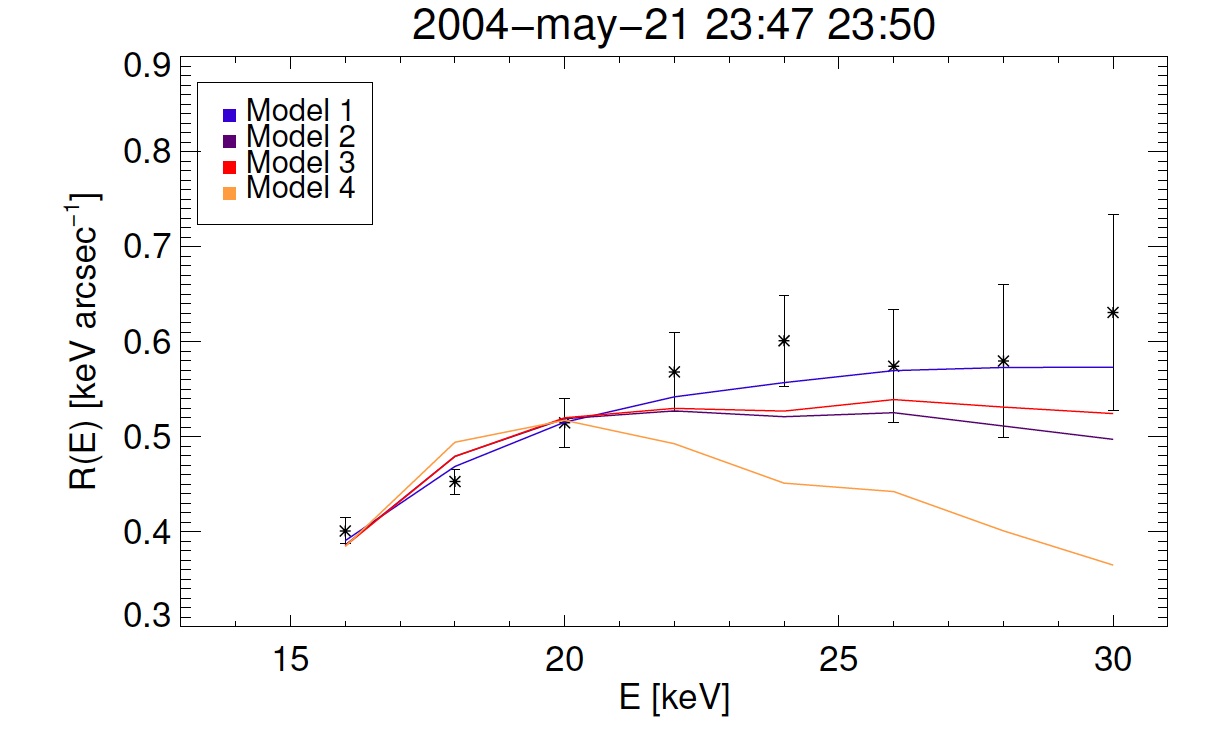}
\includegraphics[width=0.5\textwidth]{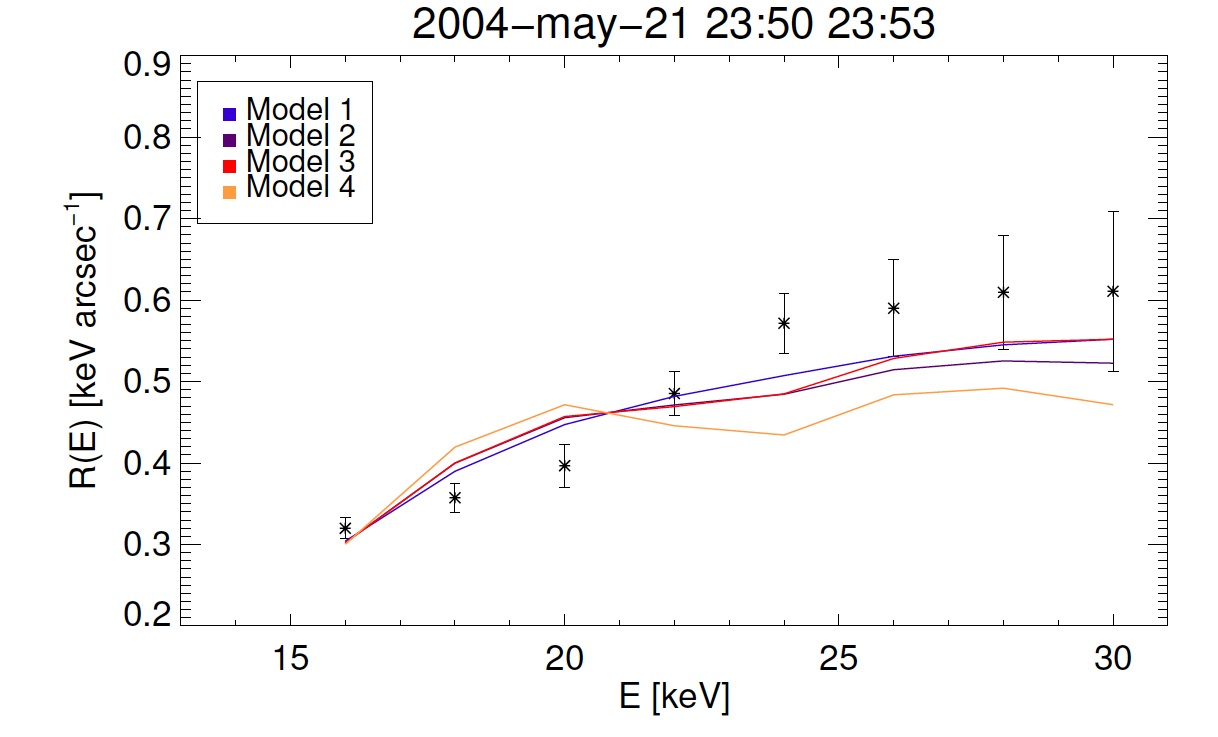}\\
\end{tabular}
\caption{15 April 2002, 16 April 2002 and 21 May 2004 events: $R(E)$ as a function of electron energy, the blue line is the best fit hot model with return currents, the purple line is the cold model with return currents, the red one 
is the model with ohmic losses, while the orange line is for the hot model without return currents .}
\label{fig:rate-position}
\end{figure}

The values of $h_s$ and $n$ are determined as described in the previous section, i.e. by fitting the empirical $R(E,s)$ values deduced from the mean electron flux spectral images and averaged along $\Delta s$, against the four models of the energy loss rate. These best-fitted values for $h_s$ and $n$ in the three events under analysis are given in Table 2 where we also provide the corresponding values of the kinetic energy associated to the return currents' velocity. Furthermore, Figure \ref{fig:rate-position} contains the empirical values of $R(E)$ used for the fitting and shows the best-fit curves corresponding to the four theoretical models. 

From these results we notice first that for the April 16 2002 event (in both considered time intervals), the $\chi^2$ values corresponding to the four models are very similar. Correspondingly, the values of the kinetic energy $E_0$ associated to the drift velocity $w_0$ obtained for this event are considerably smaller than for the other two events. 
This is probably a consequence of the fact that this event in these time intervals is in a late phase (coherently, the panels in Figure \ref{fig:rate-position} corresponding to this event show decreasing values for $R(E,s)$). For the other data sets, Model 1 systematically provides smaller $\chi^2$ values. 
Knowing the emission measure obtained from photon spectral fitting, and that $EM = Vn^2$, where $V$ is the flare volume,
we can validate the obtained density values.
Assuming that the flare has the shape of a tube, we can estimate $V$ for every time intervals starting from the electron flux maps as previously done for the area $A$. We obtain that the emission measure obtained for Model 1 has an order of magnitude in agreement with the emission measure obtained from photon spectral fitting. On the contrary, the emission measure predicted by Model 3 is two orders of magnitude lower.

\begin{center}
\begin{table}[!h]
\begin{tabular}{|c|c|c|c|c|c|}
\hline
\multicolumn{6}{ |c| }{{\bf 15$-$Apr$-$2002}} \\
\hline
 Time $[UT]$ & $EM [10^{49} cm^{-3}]$& $kT [keV]$ & $\gamma$ & $A [arcsec^2]$ & $\Delta s [arcsec]$\\
\hline
 00:03:00-00:06:00 & $0.224\pm 0.013$ & $2.03\pm 0.03$ & $8.2\pm 0.2$ & $388$ & $[-7.2, - 2.0]$\\
\hline
 00:06:00-00:09:00 & $0.412\pm 0.025$ & $1.88\pm 0.03$ & $8.1\pm 0.2$ & $460$ & $[- 7.8, -2.0]$ \\
\hline
 00:09:00-00:12:00 & $0.51\pm 0.04$ & $1.84\pm 0.03$ & $8.3\pm 0.2$ & $522$& $[- 9.2,-3.4]$ \\
\hline
\hline
\multicolumn{6}{ |c| }{{\bf 16$-$Apr$-$2002}} \\
\hline
 Time $[UT]$ & $EM [10^{49} cm^{-3}]$& $kT [keV]$ & $\gamma$ & $A [arcsec^2]$ & $\Delta s [arcsec]$\\
 \hline
 13:10:00-13:15:00 & $0.331\pm 0.022$ & $1.83\pm 0.03$ & $8.3\pm 0.1$ & $480$ & $[ \,\,1.0 , \,\,6.2]$ \\
\hline
 13:15:00-13:20:00 & $0.55\pm 0.04$ & $1.613\pm 0.021$ & $9.3\pm 0.1$ & $483$& $[\,\,1.0 , \,\,9.1]$ \\
\hline
\hline
\multicolumn{6}{ |c| }{{\bf 21$-$May$-$2004}} \\
\hline
 Time $[UT]$ & $EM [10^{49} cm^{-3}]$& $kT [keV]$ & $\gamma$ & $A [arcsec^2]$ & $\Delta s [arcsec]$\\
 \hline
 23:47:00-23:50:00 & $0.354\pm 0.027$ & $1.85\pm 0.03$ & $8.1\pm 0.1$ & $215$ & $[\,\,1.0,\, \,3.4]$\\
\hline
 23:50:00-23:53:00 & $0.62\pm 0.04$ & $1.75\pm 0.03$ & $8.5\pm 0.1$ & $216$ & $[\,\,1.0 , \,\,4.4]$\\
\hline
\end{tabular}
\caption{Main characteristics of the events under analysis. First column: time range considered; second column: emission measure determined from spectroscopy; third column: temperature determined from spectroscopy; fourth column: photon spectral index determined from spectroscopy; fifth column: flare area computed from the reference image at 14 keV; sixth column: averaging interval determined as in \cite{to12}.}
\end{table}
\end{center}

\begin{center}
\begin{table}
\begin{tabular}{|c|c|c|c|c|c|}
\hline
\multicolumn{6}{ |c| }{{\bf 15$-$Apr$-$2002}} \\
\hline
Time[UT]& Mod & $E_0[keV]$ & $n[cm^{-3}]$& $h_{s}[cm^{-5}keV^{-1}s^{-1} ]$ & $\chi^2$ \\
\hline
00:03:00-00:06:00 & Mod 1  & 
$      7.7\pm     0.2$ 	&
$   (8.0 \pm  0.8) \times10^{10}$ 	&
$	(1.6 \pm 0.3)\times10^{30}	$				&
$ 0.80$                           	\\
\hline
 & Mod 2 &
$ 2.5 \pm    0.1$					&
$ (1.81 \pm  0.16)\times10^{11}$		&
$	(4.7\pm0.8)\times10^{30}	$					&
$1.04$								\\
\hline
& Mod 3 &
$ 7.9\pm       1.9 $ &
$(7.86\pm0.12)\times10^{9} $ &
$(4.2\pm   0.5)\times 10^{30}$&
$0.82$\\
\hline
& Mod 4			& 
$0$ 									& 
$(3.4 \pm 0.3)\times10^{10}$       &
$ (9.52 \pm 1.19)\times10^{30} $						&
$3.58$\\
\hline
00:06:00-00:09:00& Mod 1  &
$       8.2\pm     0.3$&
$   (6.3 \pm  0.6)\times10^{10}$&
$	(1.04 \pm  0.21)\times10^{30}	$				&
$0.33$\\
\hline
& Mod 2&
$       3.2 \pm    0.1$&
$   (1.42 \pm  0.11)\times10^{11}$&
$	(3.3 \pm  0.5)\times10^{30}    $&
$0.55$\\
\hline
&  Mod 3 &
$ 3.5\pm      0.8 $ &
$(6.99\pm   0.12)\times10^{9} $ &
$(3.5\pm   0.4)\times 10^{30}$&
$0.43$\\
\hline
& Mod 4			& 
$0$										 & 
$   (2.30\pm  0.19)\times10^{10}$		&
$(7.2 \pm 0.8)\times 10^{30}$                            &
$2.45$\\
\hline
00:09:00-00:12:00 & Mod 1&
$       4.9\pm      1.2$&
$   (1.6\pm  0.3)\times10^{10}$&
$	(5.1 \pm 2.3)\times10^{28}  $&
$0.65$\\
\hline
& Mod 2&
$       3.2\pm     0.6$&
$   (4.5\pm  0.7)\times10^{10}$&
$   (3.45\pm 1.11)\times10^{29} $&
$0.70$\\
\hline
&  Mod 3 &
$ 0.2\pm 0.1 $ &
$(4.47\pm  0.11)\times10^{9} $ &
$(1.00\pm 0.24)\times 10^{30}$&
$0.79$\\
\hline
&Mod 4& $0$ & 
$   ( 6.2\pm  0.5)\times10^{9}$ &
$(1.4\pm0.3)\times10^{30}$						&
 $1.21$\\
 \hline
\hline
\multicolumn{6}{ |c| }{{\bf 16$-$Apr$-$2002}} \\
\hline
Time [UT] & Mod & $E_0[keV]$ & $n[cm^{-3}]$& $h_{s}[cm^{-5}keV^{-1}s^{-1} ]$ & $\chi^2$ \\
\hline
13:10:00-13:15:00 & Mod 1    	& 
$      1.8\pm     1.7$ 			&
$   (1.5 \pm  0.6) \times10^{10}$ 	&
$	(2.8 \pm  2.5)\times10^{28}	$				&
$1.22$                           	\\
\hline
 & Mod 2 &
$ 1.2 \pm    0.8$					&
$ (4.9 \pm  1.8)\times10^{10}$		&
$ (2.6 \pm  1.8)\times10^{29}	$					&
$1.18$								\\
\hline
&  Mod 3 &
$ 0.1 \pm 0.1 $ &
$(7.7\pm   0.3)\times10^{9} $ &
$(5.7\pm 3.9)\times 10^{29}$&
$1.28$\\
\hline
 & Mod 4		& 
$0$ 									& 
$(1.2 \pm 0.3)\times10^{10}$       &
$(1.3\pm 0.8)\times10^{30} $						&
$1.12$\\
\hline
13:15:00-13:20:00 & Mod 1  &
$       2.1		\pm     2.0$&
$   (1.4 \pm  0.7)\times10^{10}$&
$	(3.23 \pm  3.24)\times10^{28}$	&
$0.20$\\
\hline
 & Mod 2&
$       1.4 \pm    0.9$&
$   (4.6 \pm  1.8)\times10^{10}$&
$	(2.5 \pm 2.0)\times10^{29}    $&
$0.20$\\
\hline
&  Mod 3 &
$ 0.0\pm0.1 $ &
$(7.3\pm0.6)\times10^{9} $ &
$(6.8\pm 4.8)\times 10^{30}$&
$0.23$\\
\hline
 & Mod 4			& 
$0$										 & 
$  (9.5\pm 2.3)\times10^{9}$		&
$  (1.0 \pm 0.8) \times10^{30}$                            &
$0.30$\\
\hline
\hline
\multicolumn{6}{ |c| }{{\bf 21$-$May$-$2004}} \\
\hline
Time[UT]& Mod & $E_0[keV]$ & $n[cm^{-3}]$& $h_{s}[cm^{-5}keV^{-1}s^{-1} ]$ & $\chi^2$ \\
\hline
23:47:00-23:50:00& Mod 1  &
$       8.7\pm     0.2$&
$   (7.6\pm  0.6)\times10^{10}$&
$	(1.55 \pm  0.23)\times10^{30}	$				&
$3.85$\\
\hline
& Mod 2&
$       4.0\pm   0.1$&
$   (1.69 \pm  0.10)\times10^{11}$&
$	(5.2 \pm  0.6)\times10^{30}    $&
$12.11$\\
\hline
&  Mod 3 &
$ 5.5\pm0.9 $ &
$(6.65\pm0.08)\times10^{9} $ &
$(6.3\pm 0.5)\times 10^{30}$&
$10.43$\\
\hline
& Mod 4				& 
$0$										 & 
$   (2.23\pm  0.14)\times10^{10}$		&
$(1.16 \pm0.11)\times 10^{31}$                            &
$40.91$\\
\hline
23:50:00-23:53:00 & Mod 1     			& 
$      8.3\pm     0.2$ 			&
$   (8.3 \pm  0.6) \times10^{10}$ 	&
$	(1.78 \pm 0.24)\times10^{30}	$				&
$ 5.42$                           	\\
\hline
 & Mod 2 &
$3.7 \pm    0.1$					&
$ (1.81 \pm  0.10)\times10^{11}$		&
$	(5.7\pm0.6)\times10^{30}	$					&
$17.58$								\\
\hline
& Mod 3 &
$ 3.5\pm0.5 $ &
$(7.79\pm0.08)\times10^{9} $ &
$(1.05\pm 0.06)\times 10^{31}$&
$21.35$\\
\hline
&Mod 4		& 
$0$ 									& 
$( 2.42 \pm 0.13)\times10^{10}$       &
$ (1.21 \pm 0.11)\times10^{31} $						&
$54.12$\\
\hline
\end{tabular}
\caption{Plasma parameters obtained with the four models described above. First column: data time intervals; second column: energy loss rate models used for the fit; third column: return currents' energies obtained from the fit;  fourth column: target densities obtained from the fit; fifth column: averaged source amplitudes obtained from the fit; sixth column: $\chi^2$ values of the fit.}
\end{table}
\end{center}

\section{Conclusion}
We have considered the influence of the background electrons drift velocity
 in the accelerated electrons energy loss rate due to Coulomb collisions.
The background motion has been explained in terms of return currents that instantaneously occur in solar flares to restore
charge equilibrium.The obtained model for the energy loss rate has been compared with previous models 
by means of a validation process based on the use of averaged electron flux maps obtained from RHESSI hard X-ray imaging spectroscopy data for three different sources (April 15 2002, April 16 2002 and May 21 2004 ).\\
In all the analyzed cases, when return currents are significant to describe the observations (i.e., except in the case of the April 16 2002 data sets, where all models are comparable), the new model better fits the observed data. The main impact of this result on the theoretical picture of solar flares is concerned with the effectiveness of the emission process. Indeed the background coherent motion due to return currents tends to lower the energy loss rate of Coulomb collisions, and therefore the path of the electrons injected in the flare tends to be longer.
Accelerated electrons have thus more time to emit hard X-rays by bremsstrahlung, which increases the efficiency of the emission process.

\begin{acknowledgments}
\end{acknowledgments}


\newpage
\begin{thebibliography}{}{}

\bibitem[Alexander \& Daou(2007)]{alexanderdaou07} Alexander,~D., \& Daou,~A.~G. 2007, \apj, 666, 1268

\bibitem[Brown(1971)]{brown71} Brown,~J.~C. 1971, \solphys, 18, 489

\bibitem[Brown et al(2008)]{bretal08} Brown,~J.~C., Kasparova,~ J., Massone,~A.~M., \& Piana,~M. 2008, Astronomy and Astrophysics, 486, 1023-1029

\bibitem[Brown \& Bingham(1984)]{brownbingham84} Brown,~J.~C., \&  Bingham,~R. 1984, Astronomy and Astrophysics, 131, L11

\bibitem[Butler \& Buckingham(1962)]{bubu62} Butler, S. T. \& Buckingham, M. J. 1962, Physical Review, 126, 1

\bibitem[Emslie(1978)]{em78} Emslie,~A.~G. 1978, \apj, 224, 241

\bibitem[Emslie(1980)]{emslie80} Emslie,~A.~G. 1980, \apj, 235, 1055

\bibitem[Emslie(1981)]{emslie81} Emslie,~A.~G. 1981, \apj, 249, 817

\bibitem[Emslie, Barrett, \& Brown(2001)]{ebb01} Emslie, A.~G., Barrett, R.~K., \& Brown, J.~C. 2001, \apj, 557, 921

\bibitem[Fletcher et al.(2011)]{fle11} Fletcher, L., Dennis, 
B.~R., Hudson, H.~S., et al.\ 2011, \ssr, 159, 19 

\bibitem[Frankel(1965)]{fr65} Frankel,~N.~E. 1965, Plasma Phys., 7, 225

\bibitem[Guo et al.(2012)]{guo12} Guo,~J., Emslie,~A.~G., Kontar,~E.~P., Benvenuto,~F., Massone,~A.~M., \& Piana,~M. 2012, \apj, 755, 32

\bibitem[Hannah \& Kontar(2011)]{hannahkontar11} Hannah,~I.~G., \& Kontar,~E.~P. 2011, \aap, 529, 109
 
\bibitem[Holman(2012)]{holman2012} Holman,~G.~D. 2012, \apj, 745, 52

\bibitem[Hoyng \& Melrose(1977)]{hoyngmelrose77} Hoyng,~P., \& Melrose,~D.~B. 1977, \apj, 218, 866

\bibitem[Knight \& Sturrock(1977)]{knightsturrock77} Knight,~J.~W., \& Sturrock,~P.~A. 1977, \apj, 217, 306

\bibitem[Larosa \& Emslie(1989)]{larosaemslie89} Larosa,~T.~N., \&  Emslie,~A.~G. 1989, \solphys, 120, 343

\bibitem[Lin et al.(2002)]{linetal02} Lin,~R.~P., et~al. 2002, \solphys, 210, 3

\bibitem[Longmire(1963)]{lg63} Longmire,~C.~L. 1963, Elementary Plasma Physics (New York: Interscience)

\bibitem[Piana et al.(2007)]{pianaetal07} Piana,~M., Massone,~A.~M., Hurford,~G.~J., Prato,~M., Emslie,~A.~G., Kontar,~E.~P. and Schwartz,~R.~A. 2007, \apj, 665, 846

\bibitem[van den Oord(1990)]{oord90} van den Oord,~G.~H.~J. 1990, Astronomy and Astrophysics, 234, 496

\bibitem[McClymont \& Canfield(1986)]{mcclymontcanfield86} McClymont,~A.~N., \& Canfield,~R.~C. 1986, \apj, 305, 936

\bibitem[Rutherford(1911)]{ruth11} Rutherford, ~E. 1911, The London, Edinburgh, and Dublin Philosophical Magazine and Journal of Science, 21.125 669-688.

\bibitem[Spitzer(1962)]{sp62} Spitzer,~L.,~Jr. 1962, Physics of Fully Ionized Gases (New York: Interscience)

\bibitem[Torre et al.(2012)]{to12} 
Torre,~G.,   Pinamonti,~N.,    Emslie,~A.~G.,  Guo,~J.,   Massone,~A.~M.,   Piana,~M. 2012, \apj, 751, 129

\bibitem[Zharkova \& Gordovskyy(2005)]{zharkovagordovskyy05} Zharkova,~V.~V. \& Gordovskyy,~M. 2005, Astronomy and Astrophysics, 432, 1033

\end{thebibliography}
\end{document}